\documentclass[aps,pra,reprint,noshowpacs,superscriptaddress, nofootinbib]{revtex4-1}   
\usepackage[T1]{fontenc}	% should generally be included for better accented-word behavior
\usepackage[latin9]{inputenc}	% should generally be included for better accent behavior
\usepackage{geometry}		% for controlling page margins
\usepackage{graphicx}
\usepackage[above,below]{placeins}	% allows use of \FloatBar­rier command to force section barriers
\usepackage{times}
\usepackage{textcomp}
\begin{document}

\title{Quantitative Measures of Equity in Small Groups}
\author{Ben Archibeque}
\affiliation{Department of Physics, Kansas State University, 1228 N. 17th St., Manhattan, KS, 66506}
\author{Florian Genz}
\affiliation{Future Strategy of Teacher Education (ZuS), University of Cologne/Germany, G\"{u}rzenich Str. 27, 50667 K\"{o}ln }
\author{Maxwell Franklin}
\affiliation{Department of Physics, Swarthmore College, 500 College Avenue, Swarthmore, PA, 19081-1397}
\author{Scott Franklin}
\affiliation{School of Physics and Astronomy, RIT, 1258 Carlson Building, 54 Lomb Memorial Drive, Rochester, NY 14623}
\author{Eleanor Sayre}
\affiliation{Department of Physics, Kansas State University, 1228 N. 17th St., Manhattan, KS, 66506}

\begin{abstract}
This project investigates how to quantitatively measure equity in small student groups. We follow several student groups to operationalize how discourse may be equitable or inequitable. The groups came from a two week, pre-college program that prepares first generation and deaf/hard-of-hearing students to major in a STEM field. In the program, students focus on improving their metacognitive skills and cultural preparation for college life within a context of model building. We use three methods to measure equity. First, we look at speaking time: who talks, when, and to whom. Second, we look to segment student discourse by analyzing consistency of group speaking time. Third, we analyze group equality over time changes. We analyze these methods to see how effective they are at capturing equity in group discourse. 

\end{abstract}

\maketitle
\section{Introduction}
Group work is a fundamental part of most research-based teaching methods in physics~\cite{mcdermott_tutorials_2002,keller_research-based_2007,redish_research-based_2007,heller_cooperative_nodate}. Research on group work focuses on group effectiveness~\cite{keil_identifying_2015}, reasoning~\cite{hoehn_conceptual_2016}, and coordination~\cite{barron_achieving_2000}. These studies focus on "mutuality"~\cite{barron_achieving_2000} and "equity of voice,"~\cite{keil_identifying_2015}. Mutuality is whether individuals in a group can contribute to the group, which was measured by looking at turn-taking norms. Equity of voice is if individuals contribute to the group, as measured by speaking time. This paper will analyze each of these aspects more in depth by focusing on speaking time and how it changes over time within a group. Throughout the paper we will use equality to mean that group members are participating in the same amount and equity to mean each group member has affordances to participate.

Speaking time is readily measurable and closely related to dominance~\cite{mast_dominance_2002}. If all group members have similar dominance over the group, the group is considered equitable. The relationship between dominance and speaking time is stronger for men and increases as group size increases~\cite{mast_dominance_2002}. 

Talking time also relates to status within a group; specifically, high status students talk more frequently and present more ideas while low-status students are excluded from participation and are seen as inept~\cite{bianchini_1997}. Students who talk less may also learn less~\cite{cohen_1989}. The goal of group work is to increase learning but the group must be equitable for that learning to be consistent across all students.

Speaking patterns are unpredictable and not equitable if an individual is oppressed~\cite{gutierrez_enabling_2002}. Therefore, we suggest individuals who speak two times more than their group mates are oppressing those group mates. We chose a threshold of 200\% because it is small enough for there to be larger cases but small enough to possibly not occur at all (both seen in~\cite{keil_identifying_2015}). The severity of inequality in our analysis relates to both how many times larger the speaking time difference is and between how many group members this is shared.

As a second measure we analyze conversation equality over time, and the average overall, to understand how equity and equality may relate to one another. How we measured equality over time is discussed later.

Finally, we examine the consistency of speaking time, which shows how a group interacts over time and provides a way to segment videos. If a student begins silent and ends by being the predominant speaker within the group, this could be missed by total speaking time but is easily visible in our conversational consistency analysis, which can also distinguish if a person's speaking time is front heavy, or sporadic, which all could have different implications about a group. We do not discuss these implications in depth here; we simply use the method to see how it might distinguish group behaviors. Segmenting videos is beneficial because it provides less biased segmentation of video for various, possibly more qualitative, forms of analysis. 
% * <florian.genz@uni-koeln.de> 2017-07-03T13:01:48.994Z:
%
% > Segmenting videos is beneficial because it provides less biased segmentation of video for various, possibly more qualitative, forms of analysis. 
%
% ^.

\section{Context}

   We recorded more than 100 hours of video of students in large group (approximately twenty students) discussions and small group (approximately four students) experiments, discussions, and activities. The students are at a private, doctoral-granting technical institute in the northeast in a two week, pre-college bridge program for first generation (FG) and deaf or hard-of-hearing (DHH) individuals to prepare them for their STEM field major by reinforcing metacognitive practices. Both FG and DHH identities are often marginalized in college settings; additionally, our participants often hold multiple marginalized identities including gender, sexual, racial, and ethnic identities. In this analysis we ignore those aspects because we suspect inequality can happen in any setting. In the groups we present in this paper, no students used ASL or an interpreter as their primary mode of communication with their group mates. 
   
      We analyzed two groups: one which seemed more equitable and one which seemed less equitable. One student is in both groups. We used three different measures of equity for each group: the first is \textit{overall speaking time and distribution}, the second we call \textit{conversation consistency}, and the third we call \textit{conversation equality}.

\section{Methods}

%\subsection{Overall speaking time}

To measure speaking time we used Behavioral Observation Research Interactive Software (BORIS)~\cite{friard_boris:_2016}, an open source video and audio coding software originally designed to analyze animal behaviors. It allows for coding of states, e.g. when a student is talking, and events, e.g. when a student says a specific word. In our analysis, a student was considered speaking when they spoke loud enough to be heard (by the researcher watching the video, not by the group members) or, regardless of volume, if they occupy the floor, i.e. someone is listening to what they are saying. This addendum is necessary when the audio quality of the video is low or background noise is high. Coding for speaking time in BORIS involved selecting when an individual starts and stops speaking. We considered interjections (hmm, uhh, etc.) part of speaking time if they were preceded or followed by other words from the speaker. Pauses which were shorter than 0.3 seconds were considered talking. If the pause is longer than 0.3 seconds, the person is considered talking during that time only if they are in the middle of a sentence and are uninterrupted. 

   We describe their speaking time as the percentage of time an individual spoke compared to the difference between when the first person started talking and when the last person stopped talking. Equity in this measure is defined as similar amounts of speaking time for each group member, less than two times larger.
   
The following two methods deal with creating conversation vectors. The use of these vectors is novel to our analysis. We decided to use speaking time vectors rather than turns of talk vectors because the former is more suitable for overlapping or interjecting talk. We analyze them alongside the overall speaking time to clarify that speaking time alone doesn't wholly represent the equity of a group.
   
%   We also discuss the distribution of speaking by using timeline plots which show who spoke when. We analyze this in the context of our next measure of equity.
   
%\subsection{Conversation Consistency}
   To analyze the consistency of a group's conversation we broke an episode into equally sized time bins. Then, we analyzed each bin by creating a ``conversation vector`` for it. A conversation vector has one dimension for each group member, and the length in each direction is how long the person spoke in that bin. To make conversation vectors easier to compare across groups, we normalized them with respects to bin length\footnote{Normalizing also nullifies group silence and when people talk over one another. If the conversation vector is (0.5,0.5,0.5,0.5), this could be all individuals speaking the same amount and never talking over one another but it could also be all individuals talking over one another all the time.}. Finally, we take the dot product of two neighboring conversation vectors over the whole episode. The dot product shows how similar neighboring bins are. A dot product close to 1 means the conversation vectors were very similar while a dot product close to 0 means the conversation vectors were extremely different.  We call strings of highly consistent bins segments. All this information can be represented by what we call ``conversational consistency plots.``
   
   We chose to make bins that were a divisor of the total episode length, in seconds, because we wanted a homogenized vector length instead of having a vector at the end which was a different length. To decide how many bins to use, we made consistency plots with several different bin amounts and used the graph which best exhibited trends present in most of the produced plots; we expected this would be between 10 and 30 bins. Small bins had eclectic consistency values; large bins show no trends. We sectioned the video into 18 bins, which makes each bin approximately 100 seconds. 

%\subsection{Conversation equality}
   The final form of analysis we use helps to understand speaking time equality within each conversation vector. To do this, we used the same bins and normalized conversation vector as before but we dotted the conversation vector with the equality vector, a unit vector with equal values in each direction. Because both vectors are normalized, the range of conversation equality, for our groups of four, is between 0.5 and 1 \footnote{A least equitable conversation vector is (1,0,0,0). If dotted with the equality vector, the value is 0.5.}. Again, it is important to note that this measures equality of total speaking time and does not control for group members talking over one another.

\begin{figure}
\centering
\includegraphics[width=1\linewidth, height=4.2cm]{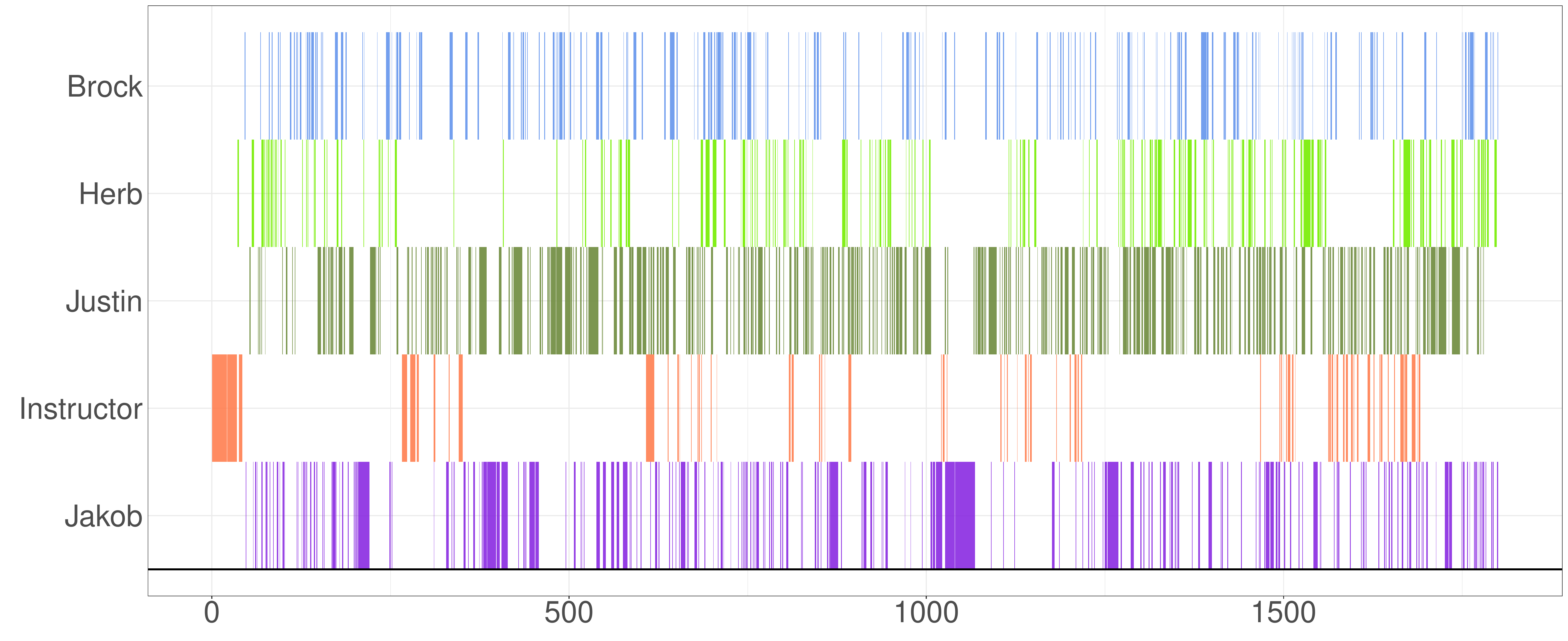}
\caption{Group 1's conversation timeline. The horizontal axis is time, in seconds. The vertical axis is the name of who is talking. \label{fig1}}
\end{figure}

\begin{figure}
\includegraphics[width=1\linewidth, height=5.1cm]{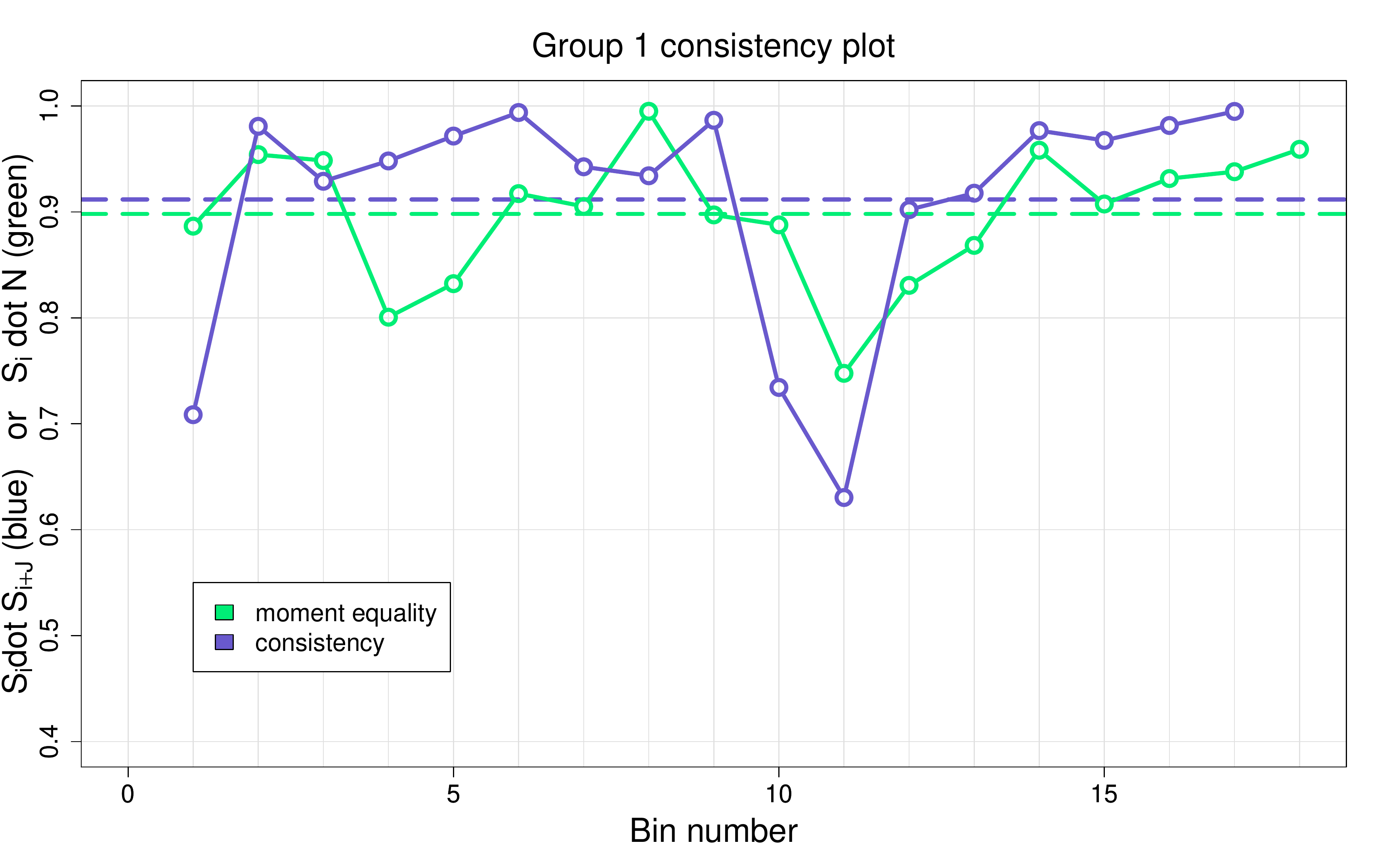}
\caption{Group 1's consistency (blue) and equality (green) plots. The dashed horizontal lines represent the mean value of equity or consistency. The y axis is the dot product value. The x axis is the conversation bin number.\label{fig2}}
\end{figure}

\section{Analysis: Group 1}
   The first group we decide to display here is from the first full day of the program. The students came in that morning and chose where they wanted to sit. The instructor had students journal their knowledge about the atmosphere for 20 minutes. Then he told the students to develop a representation of the atmosphere using the supplies on their table. 
   
   The first episode includes Brock, Herb, Jakob, and Justin. This group's timeline is shown in Fig. \ref{fig1}. Their speaking percentages are: Brock = 15\%, Herb = 10\%, Jakob = 24\%, and Justin = 35\%. The instructors spoke 10\% of the time. This group has three instances where one individual's speaking time is at least two times larger than their group mates; one of these instances is three times larger. We say this is semi-equitable. To understand this measure more fully we analyze distribution of talk to see how individuals talk throughout the activity. To do this, we begin by analyzing consistency plots then each individual within the timeline.

Figure \ref{fig2} is the consistency plot for the first episode. The first point in the graph is when the primary instructor was speaking, and two students were asking questions, and will not be analyzed here because this paper is concerned with student-student interactions. The plot indicates three segments of conversation: from bins 2 to 10, bin 11, and from bin 12 to 18.The average of their consistency values is 0.912. In the two large segments of conversation, bin 2 to 10 and bin 12 to 18, individuals speaking patterns, seen in Fig. \ref{fig1}, appear similar. It appears that bin 11, while different, did not change the group's interaction.
% * <florian.genz@uni-koeln.de> 2017-07-03T13:21:11.100Z:
% 
% > is
% legend is still wrong in Fig2 & 4
% 
% ^.
% * <florian.genz@uni-koeln.de> 2017-07-03T13:19:27.258Z:
% 
% > 2 to 10, bin 11, and from bin 12 to 18.
% would color this in Fig.2 (horizontal bars?)
% 
% ^.
% * <florian.genz@uni-koeln.de> 2017-07-03T13:08:46.383Z:
% 
% > The first point in the graph is when the primary instructor is speaking, and two students asking questions, and will not be analyzed here because this paper is concerned with student-student interactions.
% If space is rare, this can be cut out, IMO
% 
% ^.

   In the two large segments, each group member's talking distribution looks different: Justin speaks throughout the video with some periods where he talks longer. Jakob speaks often as well but talks in longer bursts than Justin. Brock and Herb have fewer, and shorter, speaking bursts than Justin and Jakob. Brock and Herb speak more sporadically, for the most part, than Justin or Jakob. Herb appears to have the largest gaps between his speaking times but they appear to grow shorter as time progresses. The large dip of consistency, is a point where Jakob talks for an extended period of time. This could be perceived as the group making room for Jakob to share his idea, however, Jakob appears to speak frequently up until that point, so this appears to be Jakob talking more, in addition to talking a lot. Therefore, we rate this group's consistency as inequitable.
% * <florian.genz@uni-koeln.de> 2017-07-03T13:23:15.297Z:
% 
% > Brock and Herb  speak more sporadically, for the most part, than Justin or Jakob.
% repetition of the sentence two before?
% 
% ^.

   With regard to this group's equality over time (see Fig. \ref{fig2}), their average equality rating is 0.898. Toward the beginning, the group's equality ratings are further from the mean but bins 12 through 15 it level at, almost exactly, the mean equality rating. None of their equality values appear to be outliers.

\section{Analysis: Group 2}
\begin{figure}
\centering
\includegraphics[width=1\linewidth, height=4.2cm]{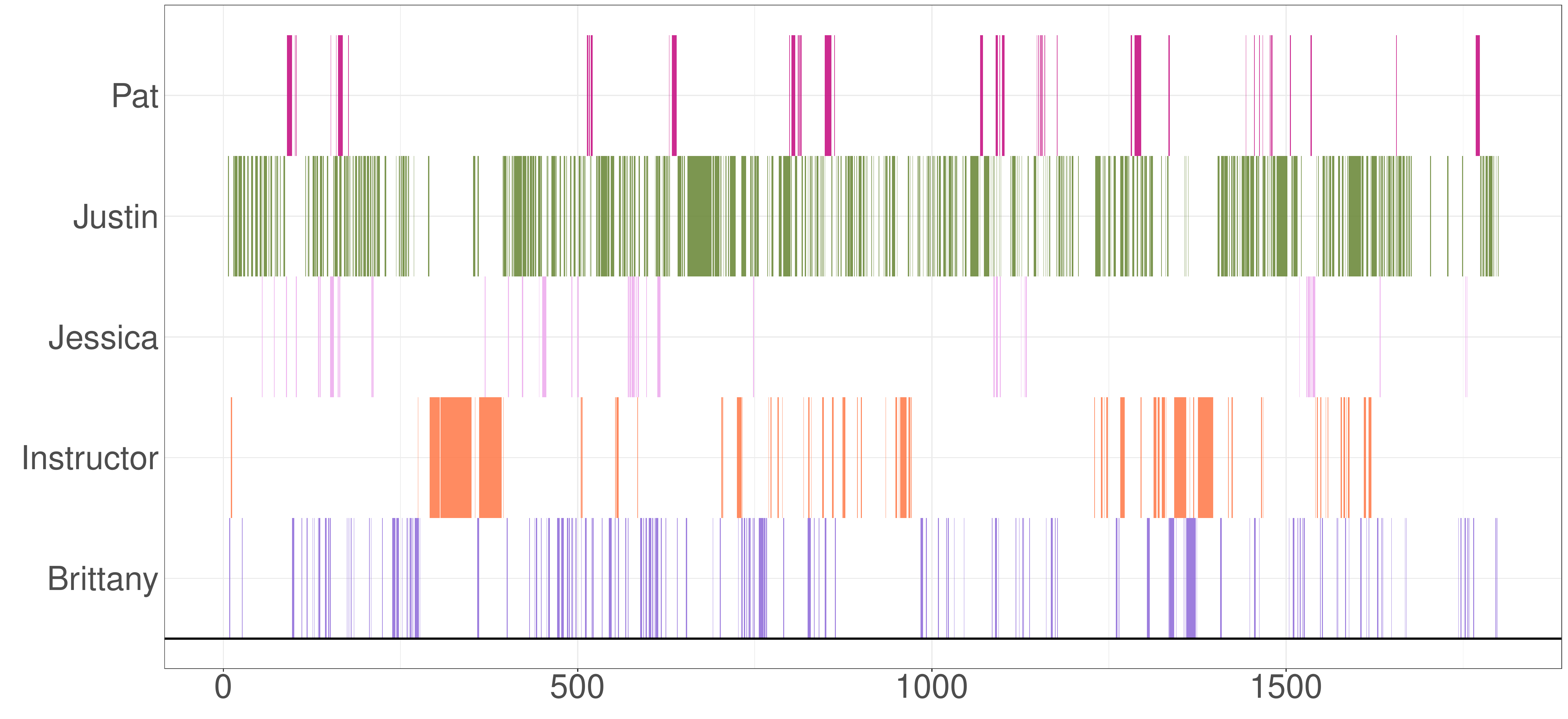}
\caption{Group 2's conversation timeline. The horizontal axis is time, in seconds. The vertical axis is the name of who is talking. \label{fig3}}
\end{figure}

\begin{figure}
\includegraphics[width=1\linewidth, height=5.1cm]{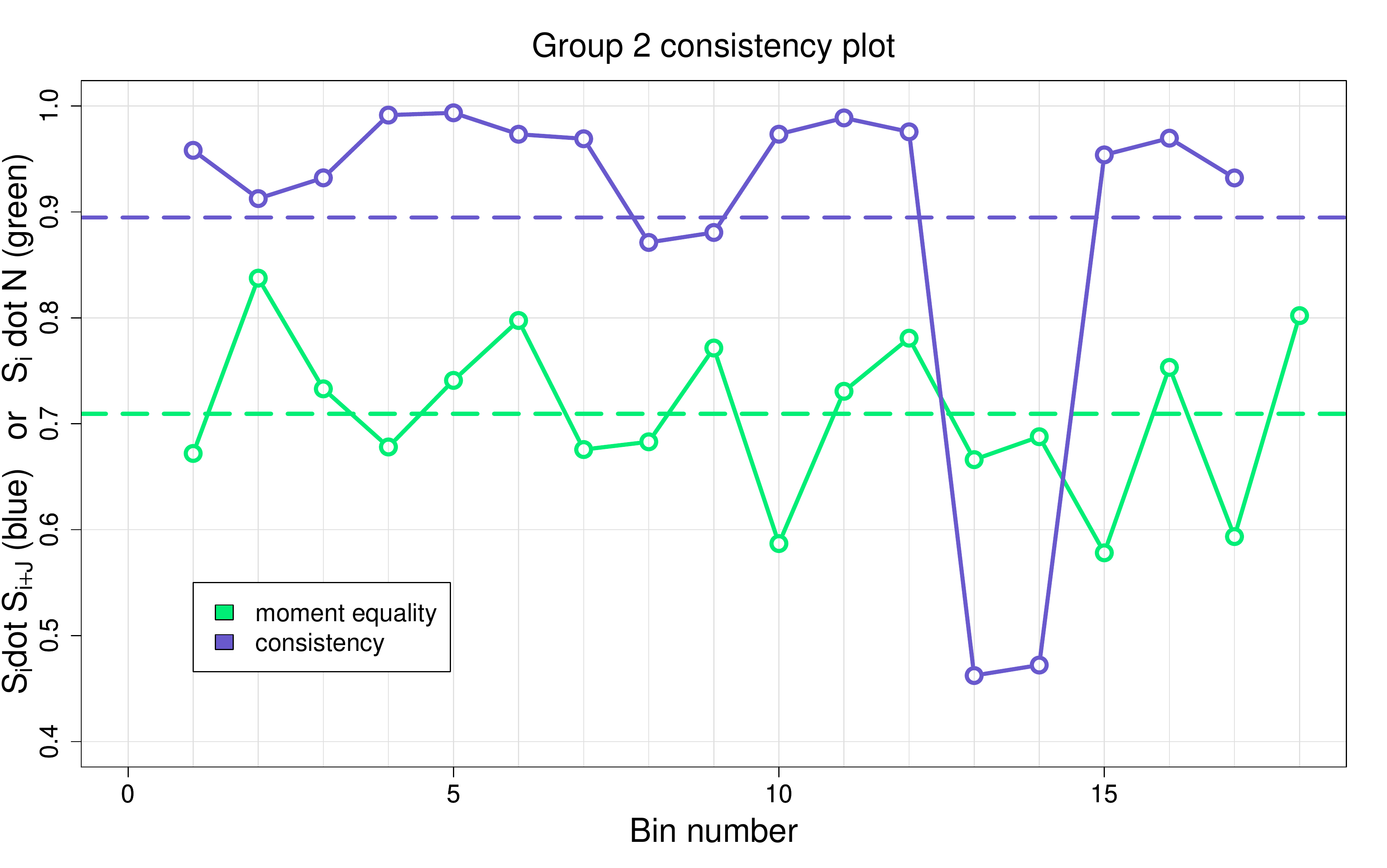}
\caption{Group 2's consistency (blue) and equality (green) plots. The dashed horizontal lines represent the mean value of equity or consistency. The y axis is the dot product value. The x axis is the conversation bin number.\label{fig4}}
\end{figure}

The next episode is from a later day when students were in new seats; the students still freely chose these seats. On this day, their task was to create an algebraic equation for the total amount of carbon dioxide that enters the atmosphere. The group members were Brittany, Jessica, Justin, and Pat. Their timeline plot (see Fig. \ref{fig3}) looks different than the previous group and so do their speaking times, which are: Brittany = 12\%, Jessica = 4\%, Justin = 38\% and Pat =5\%. The instructors' speaking time is 13\%. This group has five instances where one group member's speaking time is twice as large as their group member's speaking time. In this group, the largest difference is over nine times larger and the smallest difference is greater than two times as large. For these reasons, we say this group is inequitable with respect to total speaking time. 

    This group appears to have three episodes (See Fig. \ref{fig4}): bins 1 through 13, bin 14, and bin 15 through 18. This group's average consistency rating is 0.895, which is close to the previous group's. Justin's talking distribution looks similar to his previous group. Brittany's talking time appears to decrease slowly until bin 13, where it has a slight resurgence, after which point the group begins to talk less overall. In this bin the instructor is speaking the most and Justin is speaking considerably less than in previous bins. The consistency plot does not pertain to the instructor speaking directly but does show how it affects who is speaking in the group. If we compare the earlier time the instructor spoke with the group, there was a smaller shift in consistency, but there was still a shift. This could indicate whether the instructor impacts group dynamics or not. Returning to analyzing talking patterns, Pat talks in sporadic chunks throughout. Jessica's speaking time is small and appears to decrease, slightly, over time. We say this group is inequitable because no room is made for other group members to speak, most notably Jessica. There are a few times after bin 6 where she speaks up but there is not a large change in consistency, suggesting her group mates do not talk less to accommodate her speaking more.%Is that fair to say? Should I say something else?
% * <florian.genz@uni-koeln.de> 2017-07-03T13:25:19.620Z:
%
% > SHOULD I INCLUDE ANOTHER ONE AROUND BIN 8 AND 9
%
% ^.
    
    Figure \ref{fig4} shows this group's equality over time. The values are sporadic and, for the most part, not close to the mean, which is 0.709. This is lower than the first group. This is an example of the lower values of equality being more spread out; If one person is talking mostly, any change is substantial change. 
% * <florian.genz@uni-koeln.de> 2017-07-03T13:26:44.972Z:
% 
% > Figure \ref{fig4} shows this group's equality over time
% (in)equality?
% 
% ^.

\section{Group comparison}
 The groups' timeline and equality plots are easily distinguishable and exhibit different types of behavior. In the first group, there are more similar amounts of contribution than the second. While the distribution of speaking times was not incredibly equitable in the first group, it was more equitable than the second group.
 
The consistency plots are both similar to one another; each begins with a long string of consistent behavior and around bin 12 there is a change. This is likely a coincidence of the groups we analyzed. A group could work without any drastic changes in consistency or could work with many drastic changes, this also relates to chosen bin size. The consistency plots show times of interest for their respective group but no pattern is, necessarily, more equitable than another.
% * <florian.genz@uni-koeln.de> 2017-07-03T13:27:28.396Z:
% 
% > A group could work without any drastic changes in consistency or could work with many drastic changes, this also relates to chosen bin size. 
% What is the emphasize here? Bin size?. needs intro/transition sentence
% 
% ^.

\section{Conclusions}
   Each of these methods enlightens a different aspect of the group's behavior. Total speaking time is an effective measure of equity but fails to note if an individual's behavior over time changes drastically. For example, it misses Jessica's progressive decrease in speaking time. Conversation consistency provides insight into this missing piece of total speaking time analysis. It also allows for a more natural segmentation of a video segment for different forms of analysis. Conversation Equality is an effective, unbiased baseline to understand a groups behavior over time by quantifying how equal the group behaves. Equality appears to be consistent with inequitable groups (our unequal group was also inequitable) but appears inconsistent with equitable groups (our equal group was not highly equitable).
   
% * <florian.genz@uni-koeln.de> 2017-07-03T13:28:41.750Z:
% 
% > Equality appears to be consistent with inequitable groups (our unequal group was also inequitable) but appears inconsistent with equitable groups (our equal group was not highly equitable). 
% ???
% 
% ^.
   
%    While these methods provide a quantitative way to measure equity, they fail to regard individual differences within the group. A group member could not speak frequently but still isolate specific group members every time they speak. For this reason, we think these quantitative methods can stand alone but should also be used along side qualitative analysis to more wholly measure equity.
% * <florian.genz@uni-koeln.de> 2017-07-03T13:28:56.359Z:
% 
% > wholly
% rigorous? reliable? valid?
% 
% ^.
   
These quantitative methods can be an effective way to analyze a group. These methods scale to many group sizes and in a multitude of settings and thus are beneficial to the education community at large. To test how effective they are, for groups of various sizes and a wide array of contexts, more research is needed to validate our new measures with qualitative methods to see if they uphold as strong measures of equity or need to be refined to accommodate more types of data. We hope our quantitative measures enhance the theoretical foundation of equity in STEM labs and other contexts.	
% * <florian.genz@uni-koeln.de> 2017-07-03T13:30:49.998Z:
% 
% > To test how effective they are for groups of various sizes and a wide array of contexts, more research is needed to validate our new measures with qualitative methods to see if they uphold as strong measures of equity or need to be refined to accommodate more types of data.
% too long? grammar?
% 
% ^.

\acknowledgments{The authors are grateful to the PEER-Rochester and KSUPER groups for their thoughtful commentary and support in this paper. This material is based upon work supported by the National Science Foundation grants 1317450 and 1359262. This research was also funded by the Developing Scholars Program, the McNair Scholars Program and the BMBF ``Future Strategy of Teacher Education Cologne'' (ZuS).}

% For short, simple bibliographies, manually formatting works:
% Remember that you'll need to run pdfLaTeX twice to get the references to show (first 
% pass will insert ?? in their places).
%\begin{thebibliography}{99}

% For a longer bibliography, delete the thebibliography block above, then comment in 
% these two lines to use a .bib file with BibTeX.
\bibliographystyle{apsrev} 	% supercedes the longbibliography option, so leave commented out if you want to display article titles
%\bibstyle{apsrev}

%\bibliography{EquityReferences}

\begin{thebibliography}{0}
\expandafter\ifx\csname natexlab\endcsname\relax\def\natexlab#1{#1}\fi
\expandafter\ifx\csname bibnamefont\endcsname\relax
  \def\bibnamefont#1{#1}\fi
\expandafter\ifx\csname bibfnamefont\endcsname\relax
  \def\bibfnamefont#1{#1}\fi
\expandafter\ifx\csname citenamefont\endcsname\relax
  \def\citenamefont#1{#1}\fi
\expandafter\ifx\csname url\endcsname\relax
  \def\url#1{\texttt{#1}}\fi
\expandafter\ifx\csname urlprefix\endcsname\relax\def\urlprefix{URL }\fi
\providecommand{\bibinfo}[2]{#2}
\providecommand{\eprint}[2][]{\url{#2}}

\end{thebibliography}


\begin{thebibliography}{12}
\expandafter\ifx\csname natexlab\endcsname\relax\def\natexlab#1{#1}\fi
\expandafter\ifx\csname bibnamefont\endcsname\relax
  \def\bibnamefont#1{#1}\fi
\expandafter\ifx\csname bibfnamefont\endcsname\relax
  \def\bibfnamefont#1{#1}\fi
\expandafter\ifx\csname citenamefont\endcsname\relax
  \def\citenamefont#1{#1}\fi
\expandafter\ifx\csname url\endcsname\relax
  \def\url#1{\texttt{#1}}\fi
\expandafter\ifx\csname urlprefix\endcsname\relax\def\urlprefix{URL }\fi
\providecommand{\bibinfo}[2]{#2}
\providecommand{\eprint}[2][]{\url{#2}}

\bibitem[{\citenamefont{McDermott and
  Shaffer}(2002)}]{mcdermott_tutorials_2002}
\bibinfo{author}{\bibfnamefont{L.~C.} \bibnamefont{McDermott}}
  \bibnamefont{and} \bibinfo{author}{\bibfnamefont{P.~S.}
  \bibnamefont{Shaffer}}, \emph{\bibinfo{title}{Tutorials in {Introductory}
  {Physics}}} (\bibinfo{publisher}{Prentice Hall}, \bibinfo{year}{2002}), ISBN
  \bibinfo{isbn}{978-0-13-065364-2}.

\bibitem[{\citenamefont{Keller et~al.}(2007)\citenamefont{Keller, Finkelstein,
  Perkins, Pollock, Turpen, Dubson, Hsu, Henderson, and
  McCullough}}]{keller_research-based_2007}
\bibinfo{author}{\bibfnamefont{C.}~\bibnamefont{Keller}},
  \bibinfo{author}{\bibfnamefont{N.}~\bibnamefont{Finkelstein}},
  \bibinfo{author}{\bibfnamefont{K.}~\bibnamefont{Perkins}},
  \bibinfo{author}{\bibfnamefont{S.}~\bibnamefont{Pollock}},
  \bibinfo{author}{\bibfnamefont{C.}~\bibnamefont{Turpen}},
  \bibinfo{author}{\bibfnamefont{M.}~\bibnamefont{Dubson}},
  \bibinfo{author}{\bibfnamefont{L.}~\bibnamefont{Hsu}},
  \bibinfo{author}{\bibfnamefont{C.}~\bibnamefont{Henderson}},
  \bibnamefont{and}
  \bibinfo{author}{\bibfnamefont{L.}~\bibnamefont{McCullough}}
  (\bibinfo{publisher}{AIP}, \bibinfo{year}{2007}), pp.
  \bibinfo{pages}{128--131}.

\bibitem[{\citenamefont{Redish and Cooney}(2007)}]{redish_research-based_2007}
\bibinfo{editor}{\bibfnamefont{E.~F.} \bibnamefont{Redish}} \bibnamefont{and}
  \bibinfo{editor}{\bibfnamefont{P.}~\bibnamefont{Cooney}}, eds.,
  \emph{\bibinfo{title}{Research-{Based} {Reform} of {University} {Physics}}},
  vol.~\bibinfo{volume}{1} of \emph{\bibinfo{series}{Reviews in {PER}}}
  (\bibinfo{year}{2007}).

\bibitem[{\citenamefont{Heller and Heller}(2010)}]{heller_cooperative_nodate}
\bibinfo{author}{\bibfnamefont{K.}~\bibnamefont{Heller}} \bibnamefont{and}
  \bibinfo{author}{\bibfnamefont{P.}~\bibnamefont{Heller}},
  \emph{\bibinfo{title}{Cooperative {Problem} {Solving} in {Physics}: {A}
  {User}'s {Manual}}} (\bibinfo{year}{2010}).

\bibitem[{\citenamefont{Keil et~al.}(2015)\citenamefont{Keil, Stober, Quinty,
  Molloy, and Hooker}}]{keil_identifying_2015}
\bibinfo{author}{\bibfnamefont{J.}~\bibnamefont{Keil}},
  \bibinfo{author}{\bibfnamefont{R.}~\bibnamefont{Stober}},
  \bibinfo{author}{\bibfnamefont{E.}~\bibnamefont{Quinty}},
  \bibinfo{author}{\bibfnamefont{B.}~\bibnamefont{Molloy}}, \bibnamefont{and}
  \bibinfo{author}{\bibfnamefont{N.}~\bibnamefont{Hooker}}
  (\bibinfo{publisher}{American Association of Physics Teachers},
  \bibinfo{year}{2015}), pp. \bibinfo{pages}{163--166}.

\bibitem[{\citenamefont{Hoehn et~al.}(2016)\citenamefont{Hoehn, Finkelstein,
  and Gupta}}]{hoehn_conceptual_2016}
\bibinfo{author}{\bibfnamefont{J.~R.} \bibnamefont{Hoehn}},
  \bibinfo{author}{\bibfnamefont{N.~D.} \bibnamefont{Finkelstein}},
  \bibnamefont{and} \bibinfo{author}{\bibfnamefont{A.}~\bibnamefont{Gupta}}
  (\bibinfo{publisher}{American Association of Physics Teachers},
  \bibinfo{year}{2016}), pp. \bibinfo{pages}{152--155}.

\bibitem[{\citenamefont{Barron}(2000)}]{barron_achieving_2000}
\bibinfo{author}{\bibfnamefont{B.}~\bibnamefont{Barron}},
  \bibinfo{journal}{Journal of the Learning Sciences}
  \textbf{\bibinfo{volume}{9}}, \bibinfo{pages}{403} (\bibinfo{year}{2000}),
  ISSN \bibinfo{issn}{1050-8406}.

\bibitem[{\citenamefont{Mast}(2002)}]{mast_dominance_2002}
\bibinfo{author}{\bibfnamefont{M.~S.} \bibnamefont{Mast}},
  \bibinfo{journal}{Human Communication Research}
  \textbf{\bibinfo{volume}{28}}, \bibinfo{pages}{420} (\bibinfo{year}{2002}),
  ISSN \bibinfo{issn}{1468-2958}.

\bibitem[{\citenamefont{Bianchini}(1997)}]{bianchini_1997}
\bibinfo{author}{\bibfnamefont{J.~A.} \bibnamefont{Bianchini}},
  \bibinfo{journal}{J. Res. Sci. Teach.} \textbf{\bibinfo{volume}{34}},
  \bibinfo{pages}{1039} (\bibinfo{year}{1997}), ISSN \bibinfo{issn}{1098-2736}.

\bibitem[{\citenamefont{Cohen et~al.}(1989)\citenamefont{Cohen, Lotan, and
  Leechor}}]{cohen_1989}
\bibinfo{author}{\bibfnamefont{E.~G.} \bibnamefont{Cohen}},
  \bibinfo{author}{\bibfnamefont{R.~A.} \bibnamefont{Lotan}}, \bibnamefont{and}
  \bibinfo{author}{\bibfnamefont{C.}~\bibnamefont{Leechor}},
  \bibinfo{journal}{Sociology of Education} \textbf{\bibinfo{volume}{62}},
  \bibinfo{pages}{75} (\bibinfo{year}{1989}), ISSN \bibinfo{issn}{0038-0407}.

\bibitem[{\citenamefont{Gutierrez}(2002)}]{gutierrez_enabling_2002}
\bibinfo{author}{\bibfnamefont{R.}~\bibnamefont{Gutierrez}},
  \bibinfo{journal}{Mathematical Thinking and Learning}
  \textbf{\bibinfo{volume}{4}}, \bibinfo{pages}{145} (\bibinfo{year}{2002}),
  ISSN \bibinfo{issn}{1098-6065}.

\bibitem[{\citenamefont{Friard and Gamba}(2016)}]{friard_boris:_2016}
\bibinfo{author}{\bibfnamefont{O.}~\bibnamefont{Friard}} \bibnamefont{and}
  \bibinfo{author}{\bibfnamefont{M.}~\bibnamefont{Gamba}},
  \bibinfo{journal}{Methods Ecol Evol} \textbf{\bibinfo{volume}{7}},
  \bibinfo{pages}{1325} (\bibinfo{year}{2016}), ISSN \bibinfo{issn}{2041-210X}.

\end{thebibliography}
%\end{thebibliography}

\end{document}